\documentclass[apl,preprint]{revtex4}
\usepackage{graphicx}
\usepackage{dcolumn}
\usepackage{amssymb}
\usepackage{amsmath}
\usepackage{epsfig}
\usepackage{stmaryrd}
\usepackage{color}
\usepackage{siunitx}
\usepackage{epstopdf}
\usepackage{bm}

\begin{document}

\title{Analysis of contact stiffness in Ultrasound Atomic Force Microscopy: Three-dimensional time-dependent ultrasound modeling}
\author{D. Piras} 
\author{H. Sadeghian}
\email{hamed.sadeghianmarnani@tno.nl}
\affiliation{Netherlands Organization for Applied Scientific Research, TNO, 2628 CK Delft, The Netherlands}

\date{\today}

\begin{abstract}
Ultrasound Atomic Force Microscopy (US-AFM) has been used for subsurface imaging of nanostructures. The contact stiffness variations have been suggested as the origin of the image contrast. Therefore, to analyze the image contrast, the local changes in the contact stiffness due to the presence of subsurface features should be calculated. So far, only static simulations have been conducted to analyze the local changes in the contact stiffness and, consequently, the contrast in US-AFM. Such a static approach does not fully represent the real US-AFM experiment, where an ultrasound wave is launched either into the sample or at the tip, which modulates the contact stiffness. This is a time-dependent nonlinear dynamic problem rather than a static and stationary one. 
This letter presents dynamic 3D ultrasound analysis of contact stiffness in US-AFM (in contrast to static analysis) to realistically predict the changes in contact stiffness and thus the changes in the subsurface image contrast. 
 The modulation frequency also influences the contact stiffness variations and, thus, the image contrast. The three-dimensional time-dependent ultrasound analysis will greatly aid in the contrast optimization of subsurface nano-imaging with US-AFM.
\end{abstract} 


\maketitle

Atomic Force Microscopy (AFM) was initially developed for topographic imaging \cite{Binning1986, :/content/aip/journal/rsi/86/11/10.1063/1.4936270}. In AFM, a vibrating cantilever with a sharp tip scans a sample surface, which in turn, influences the deflection of the same cantilever. The motion of the cantilever is measured with the use of the optical beam deflection method \cite{Putman1992, Herfst2014104} and gives a high resolution image of the surface \cite{Turner2001}. 
In dynamic mode AFM, the tip-sample interaction influences the vibration mode of the cantilever \cite{Rabe1998}. The response of the cantilever specifically depends on the tip-sample interaction stiffness, i.e., the so-called contact stiffness. The local variations in contact stiffness change the contact resonance frequency of the cantilever and its vibration mode \cite{Parlak2013}.

The further combination with ultrasonic excitation resulted in a variety of methods like UFM \cite{Yamanaka1994178}, AFAM \cite{Rabe2000430, Striegler20111405}, UAFM \cite{Rabe1998}, HFM \cite{Kimura2013}. We simply refer to Ultrasound-AFM (US-AFM) to indicate the common ultrasonic excitation of the different AFM schemes mentioned above. In US-AFM either the tip or the sample is excited with an ultrasonic wave. At ultrasound frequencies (tens of MHz), the cantilever is effectively stiffened, and its stiffness can be tuned to match the stiffness of the contact, improving the image contrast \cite{Rabe2000430, 0957-4484-23-21-215703,  Parlak2013}. With US-AFM, the possibility of imaging objects below the surface of a sample has been shown \cite{Rabe2000430, Striegler20111405, Kimura2013, Passeri20102769}. Such subsurface imaging capabilities are of great interest in several fields, such as semiconductors \cite{Su2011}, life sciences \cite{Shi2012}, and measurements of local mechanical properties \cite{Burnham1989}. \\
To analyze and enhance the image contrast in US-AFM or to extract quantitative material properties, the local changes in contact stiffness should be determined \cite{Parlak2008, Striegler20111405}. Static and stationary simulations have been performed \cite{Parlak2013} to predict the changes in the contact stiffness. However, static simulations do not fully represent the nonlinear dynamic situation of US-AFM since the effect of ultrasound waves on the contact stiffness is a time-dependent, dynamic problem.
The dynamic behavior of the motion of the cantilever is influenced by the tip-sample interaction, specifically the contact stiffness, and exhibits a different resonance frequency when the tip is probing on top or far away from a subsurface feature \cite{Kimura2013}. This mechanism is responsible for the subsurface imaging contrast and can be evaluated by measuring the changes in contact resonance.

To accurately analyze the contact stiffness in US-AFM, we performed three-dimensional time-dependent ultrasound calculations using the Finite Element Method (FEM), which better represents the actual experimental conditions. Before addressing the time-dependent ultrasound calculations, we first recall the basics of contact theory and describe the FEM simulations for the static stationary case. This later allows a comparison with the time-dependent ultrasound results.

\begin{figure}\includegraphics[width=12cm]{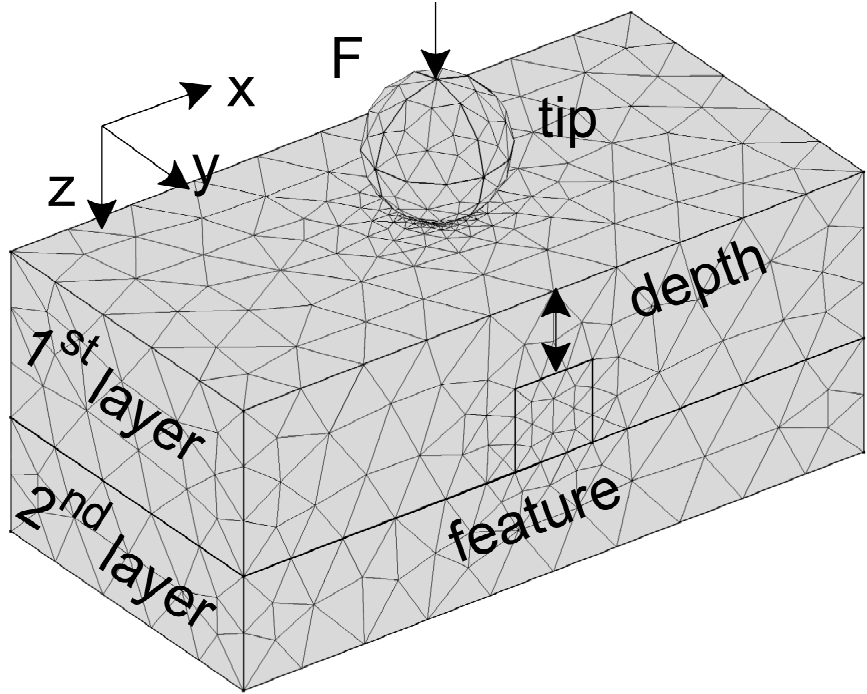}
\caption{Schematic illustration of the geometry used for FEM calculation. The origin of the system of reference coincides with the tip-first layer contact point. The x-axis coincides with the feature (or inclusion) scan direction.}
\label{fig_scheme}
\end{figure}

An implementation of a three-dimensional tip-sample contact problem using FEM has been described in \cite{Parlak2008} to estimate the effect of a scanning tip on the contact stiffness in static condition primarily for cavity-like structures (voids). This approach has been reported to be extremely time consuming due to the required fine discretization at the contact area. 
We followed the same contact approach, and to reduce the computation time and the required memory, we also implemented a semi-analytical approach based on Hertzian contact theory \cite{Hertz1896, Hanaor2015}.\\
For a spherical tip (radius $R$) pressing with a constant force $F$ on a semi-infinite and homogeneous medium in the absence of dissipative effects, the contact radius and the local stress distribution are $R_{c}=\sqrt[3]{2R\left(\frac{3}{8}\right)F\left(M_{tip}+M_{1^{st} layer}\right)}$ and $P_{c}(x,y)=\frac{3F}{2\pi R_{c}^{2}}\sqrt{1-\left(\frac{{x^{2}}+{y^{2}}}{R_{c}^{2}}\right)}$. 
In these expressions, (x,y) are the coordinates of the points on the surface of the $1$$^{st}$ layer (Fig.~\ref{fig_scheme}), $M_{tip}=\frac{E_{tip}}{\left(1-\nu _{tip}^{2}\right)}$ and $M_{1^{st} layer}=\frac{E_{1^{st} layer}}{\left(1-\nu _{1^{st} layer}^{2}\right)}$ are the reduced Young's moduli of the tip and first layer, respectively ($\nu$ is the Poisson's ratio, and $E$ is the Young's modulus). In the semi-analytical FEM approach the analytical expression of stress distribution $P_{c}(x,y)$ is imposed as a distributed load on the contact area defined by $R_{c}$. 

The output of the FEM analysis (both the contact and the semi-analytical implementation) is the indentation depth $\delta$ of the tip into the sample. 
The indentation depends on the sample structure, on the embedded finite size feature and on its position relative to the tip. The indentation is used to make a new estimate \cite{Parlak2008, Konter2006} of the reduced Young's modulus of the first layer $\left< M_{1^{st} layer} \right>=\sqrt{\frac{(3F/2\delta)^{3}}{6FR}}$. Finally, the contact stiffness $k$$^{*}$ can be calculated as $k^{*}=\sqrt[3]{6FR{\left< E^{*} \right>}^{2}}$, where $\frac{1}{\left< E^{*} \right>}=\frac{1}{M_{tip}}+\frac{1}{\left< M_{1^{st} layer} \right>}$ is the reciprocal of the effective Young's modulus.

\begin{figure}\includegraphics[width=12cm]{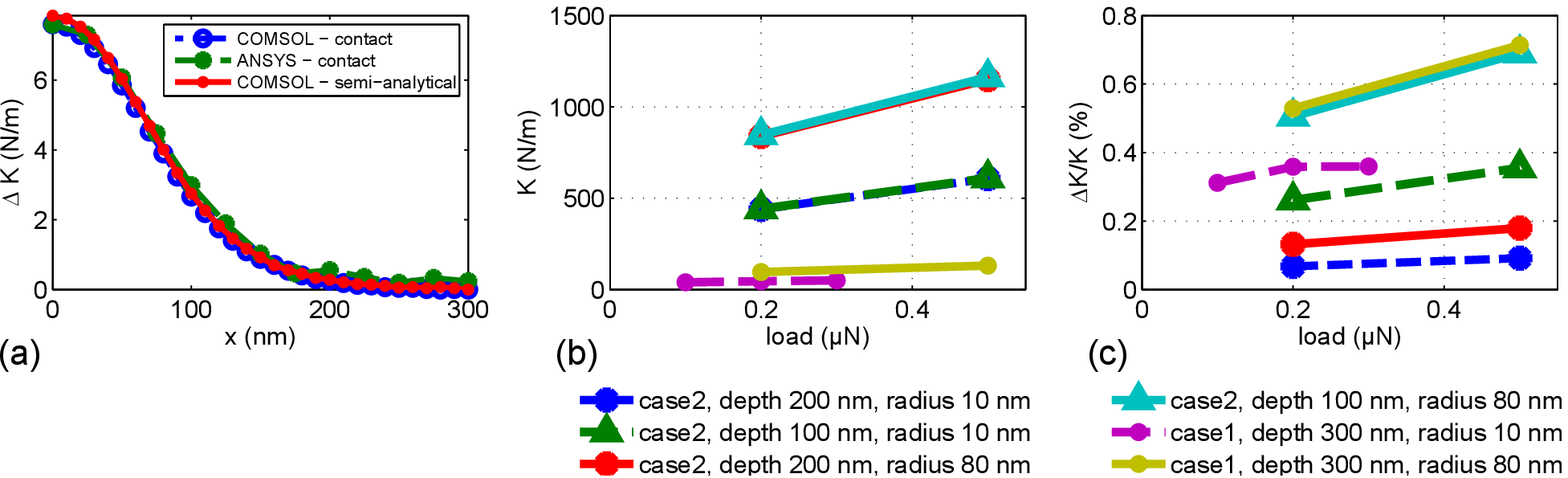}
\caption{(a) Static variations of contact stiffness versus the position of the subsurface feature with respect to the AFM tip (Table~\ref{tableGeometryproperties}. For case 2, depth$_{incl}$ is \SI{100}{nm}, and the applied force is \SI{0.5}{\micro N}): COMSOL-semi-analytical indicates the solution with the outlined procedure where the system is loaded with the Hertzian-derived contact stress distribution, ANSYS-contact and COMSOL-contact indicate the solutions with the full 3D contact procedure. 
(b) Static baseline contact stiffness (K) and (c) contact stiffness variation normalized to the baseline contact stiffness ($\Delta$K/K) for the use cases in Table~\ref{tableGeometryproperties}. In the legend, $depth$ indicates the depth of the inclusion, and $radius$ indicates the radius of the tip. The colored lines in 
(b) and (c) are given only as guides for the eye and do not imply linear interpolation.}
\label{fig_static_cases}
\end{figure}

Before implementing the ultrasound (time-dependent) wave excitation, the semi-analytical approach has been verified in the static stationary case. We show the verification for one case of a rigid inclusion in a rigid matrix (Table~\ref{tableGeometryproperties}, case 2, depth$_{incl}$ \SI{100}{nm}, force \SI{0.5}{\micro N}). The feature is moved from right below the contact in steps of \SI{10}{nm}, and the contact stiffness at the tip-sample contact location is extracted at each scan step. Fig.~\ref{fig_static_cases}(a) shows that there is very good agreement between the outlined semi-analytical procedure (calculated in COMSOL) and the 3D full contact model (calculated in both COMSOL and ANSYS). All three simulations show the same contact stiffness variation $\Delta$K of about \SI{7.5}{N/m} along the feature scan with a slight overestimation of the baseline contact stiffness for the semi-analytical procedure (1.8 \% and 1.7 \% with respect to the ANSYS contact model and COMSOL contact model, respectively) with the advantage of a substantial reduction in computation time from about 25 hours to 25 minutes for a complete scan (on an Intel Xeon 2 3.46 GHz workstation). 

\begin{table}
\caption{\label{tableGeometryproperties}Geometry used in FEM simulations. Case 1 is representative of a rigid inclusion in a soft matrix, while case 2 is representative of a rigid inclusion in a rigid matrix. The size of the layers and of the inclusion in the $z$ and $x$ directions (Fig.~\ref{fig_scheme}) are indicated by $h$ and $w$ respectively.}
\begin{ruledtabular}
\begin{tabular}{lllcccc}
case&part&material&radius&h$\times$w&depth$_{incl}$&force\\
$$&$$&$$&(nm)&(nm$\times$nm)&(nm)&(\SI{}{\micro N})\\
 \hline
 1&tip&Si&10\footnote{applied forces 0.1, 0.2 and \SI{0.3}{\micro N}}; 80&$$&$$&0.2; 0.5\\
$$&1$^{st}$ layer&PMMA&$$&$400\times800$&300&$$\\
$$&feature&Al&$$&$100\times100$&$$&$$\\
$$&2$^{nd}$ layer&Si& &$135\times800$&$$&$$\\
 \hline
2&tip&C&10; 80&$$&$$&0.2; 0.5\\
$$&1$^{st}$ layer&SiO$_{2}$&$$&$200\times800$&100&$$\\
$$&$$&$$&$$&$300\times800$&200&$$\\
$$&feature&Si&$$&$100\times100$&$$&$$\\
$$&2$^{nd}$ layer&Si&$$&$135\times800$&$$&$$\\
\end{tabular}
\end{ruledtabular}
\end{table}

The semi-analytical model has been used for all the cases listed in Table~\ref{tableGeometryproperties}. Fig~\ref{fig_static_cases}(b) shows the baseline contact stiffness (K), which is the contact stiffness when the buried feature is far from the tip. 
Fig.~\ref{fig_static_cases}(c) shows the contact stiffness variation normalized to the baseline contact stiffness ($\Delta$K/K), which is an estimation of the static contrast.

This approach allows for fast evaluations of the behaviors of certain material and load condition combinations. Fig~\ref{fig_static_cases}(b) shows that the baseline contact stiffness is largely dependent on the size of the scanned tip and the applied force.
Fig.~\ref{fig_static_cases}(c) shows that the contrast decreases when using a smaller tip (compare solid cyan and dashed green lines) and decreases further if the depth of the feature is increased (compare solid cyan and solid red lines).


The static approach gives a reference for the contact stiffness values reached in the stationary state. However, in current US-AFM techniques, the sample is subjected to an ultrasound wave excitation. The wave excitation is based on modulating a carrier frequency, $f_{c}$, with a modulation frequency, $f_{m}$, which is equal or close to the contact resonance frequency of the cantilever. Therefore, the acoustic problem must be addressed with a time-dependent approach. The semi-analytical procedure discussed above has been introduced precisely for the purpose of making the dynamic approach feasible in terms of simulation time and available memory. 
In this section, we refer to case 1 in Table~\ref{tableGeometryproperties}. The silicon layer substrate is extended in depth to \SI{550}{nm} to give more space to the ultrasound waves to propagate before reaching the $2^{nd}$ layer-$1^{st}$ layer interface.
In a bottom excitation approach, the tip is at a fixed position on top of the sample, and a prescribed displacement is delivered at the bottom surface of the silicon substrate ($2^{nd}$ layer) to simulate an ultrasound plane wave excitation. 

An eigenfrequency analysis of the sample has been performed in COMSOL. No external loads are applied, but the sample is constrained on the top surface because of the presence of the tip. The analysis shows that the first eigenmode is a compressional mode at \SI{67}{MHz}, further (flexural) modes are all above \SI{1}{GHz}.

In the time dependent simulations, the prescribed displacement is written as $\delta(t)=\delta_{0}\sin(2\pi f_\mathrm{c}t)\times \sin(2\pi f_\mathrm{m}t)$, where $\delta_{0}$=\SI{1}{nm}, $f_{c}$=\SI{67}{MHz} is the carrier frequency and $f_{m}$=[2.20, 2.21, 2.22, 2.23]~\SI{}{MHz} is the modulation frequency. In these conditions small differences in the excitation frequency are expected to provide significant differences in the behavior of the system to ultrasound propagation. The modulation frequencies are chosen in the MHz range for simulation time purposes only. In fact, in subsurface US-AFM the modulation frequency is usually selected in the vicinity of the cantilever-sample contact resonance. For a first mode of few hundreds \SI{}{kHz} it is interesting to inspect contact frequency shifts of the order of few \SI{}{Hz}. However under these conditions the simulation time scale would be exceptionally long before any difference due to modulation is appreciable. Since the aim of the paper is to evaluate the influence of ultrasound modulation on the tip-sample contact stiffness, and not on the cantilever resonance shift, the modulation frequency is chosen in the more convenient \SI{}{MHz} range. Each simulations had a total duration of \SI{2.5}{\micro s} corresponding to more than 100 carrier frequency periods and about 10 modulation frequency periods.

The tip is ensured to be always in contact with the sample, and for simplicity, the nonlinear effects of the tip approaching the sample (for example, Van der Waals forces) are neglected. To ensure this condition throughout the entire simulation time, the tip is pressed on the PMMA layer. Based on the preliminary static simulations, a \SI{0.2}{\micro N} pre-load gives a static contact deformation of the PMMA layer of \SI{3.1}{nm}. This pre-load condition ensures that the $\delta(t)$ displacement excitation never causes a detachment between the tip and the sample. 

\begin{figure}
\centering
\includegraphics[width=12cm]{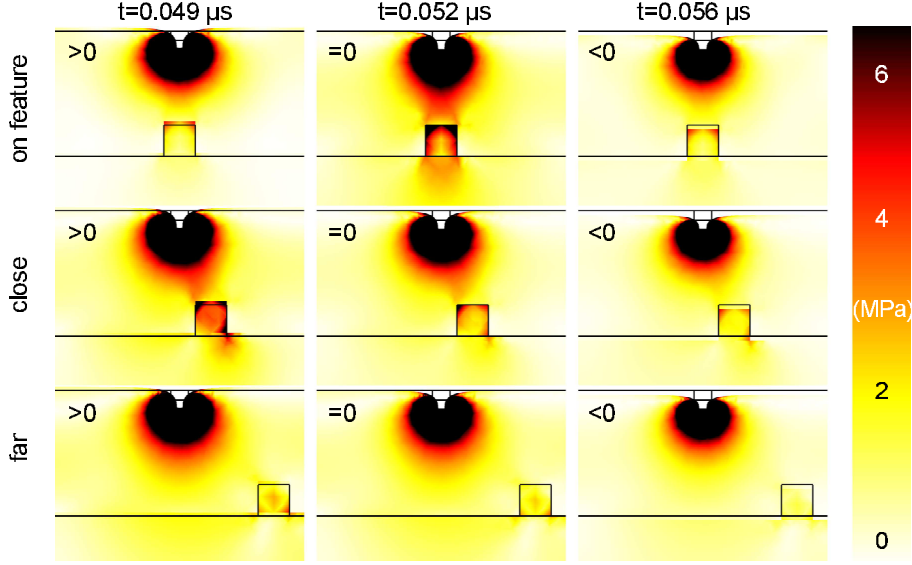}
\caption{Cross sectional stress field for different time steps (horizontally) and different positions of the feature with respect to the contact area (vertically). At \SI{0.049}{\micro s}, the substrate is compressed. At \SI{0.052}{\micro s}, the excitation is zero, and at \SI{0.056}{\micro s}, the substrate is under tensile excitation. The non-deformed top surface of the sample and the top layer/substrate and substrate/feature interfaces are marked with black solid lines. 
 The distance between the deformed free-surface and the non-deformed line, $\varepsilon$, is indicated as positive when the deformed free-surface exceeds the level of the non-deformed line.}
\label{fig_stressfield}
\end{figure}

Since the wavelengths in both the PMMA (\SI{21}{\micro m}) and silicon (\SI{110}{\micro m}) are much larger than the actual medium thickness, there is no signature of wave scattering propagation. The entire medium moves upwards/downwards following the displacement excitation, while the tip is fixed. 
Fig.~\ref{fig_stressfield} shows the cross-sectional stress distribution in the case of \SI{2.22}{MHz} modulation frequency for a few scan steps (vertically) and time steps (horizontally). 
In all the frames of Fig.~\ref{fig_stressfield}, the black solid line at the top of each frame indicates the non-deformed top surface of the sample.
At \SI{0.049}{\micro s}, the substrate is compressed: the medium moves upwards, the tip acts as a fixed boundary, the local stress extension at the contact increases and the free surface of the sample exceeds the non-deformed top surface line. 
At \SI{0.056}{\micro s}, the substrate is under tensile excitation: the medium moves downwards, the local stress extension at the contact is minimum and the free surface of the sample is below the non-deformed line. 
At \SI{0.052}{\micro s}, the excitation is approximately zero: the free surface of the sample coincides with the non-deformed line.
The closer the feature to the tip, the more evident the way in which the stress distribution is distorted by the presence of the feature.
The distance between the deformed free-surface and the non-deformed line, $\varepsilon$, is the relative displacement between the PMMA surface with respect to the fixed indentation of the tip. From each time frame and for each scan position, the value of $\varepsilon$ is estimated and is used to calculate the contact stiffness. 

\begin{figure}
\centering
\includegraphics[width=12cm]{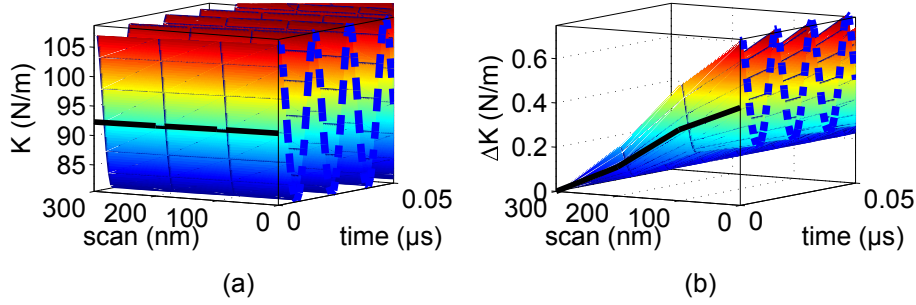}
\caption{(a) Contact stiffness and (b) variation of contact stiffness at $f_\mathrm{c}-f_\mathrm{m}$ ($f_\mathrm{m}$=\SI{2.22}{MHz}) as functions of the excitation time and feature scan position. The stationary scan contact stiffness profile is always plotted as a solid black line for reference. The dash-dot blue lines highlight the contact stiffness time profile on-feature.}
\label{fig_timedependent}
\end{figure}

Because of the modulated excitation shape, the ultrasound wave has a center frequency at $f_\mathrm{c}-f_\mathrm{m}$. As a result, the time varying indentation of the tip and the contact stiffness also have the largest spectral component, thus maximum sensitivity, at the frequency $f_\mathrm{c}-f_\mathrm{m}$. For this reason a pure harmonic centered at the frequency $f_\mathrm{c}-f_\mathrm{m}$ is used to fit the contact stiffness time trace. The fitted pure harmonic is plotted in Fig.~\ref{fig_timedependent}(a) for the modulation frequency of \SI{2.22}{MHz} for a few periods (time axis), for each scan position of the feature relative to the tip (scan axis).


At each time step, the effects of the ultrasound excitation on the contact stiffness are far larger than the effects of the feature position relative to the tip. For this reason, the off-feature contact stiffness values (at \SI{300}{nm} scan) are subtracted, and the obtained contact stiffness variations are shown in Fig.~\ref{fig_timedependent}(b). 
The comparison with respect to the stationary static case (solid black lines in a-b) shows that the use of dynamic excitation can induce increased contact stiffness on-feature at the frequency of the maximum energy content $f_\mathrm{c}-f_\mathrm{m}$. 

\begin{figure}
\centering
\includegraphics[width=12cm]{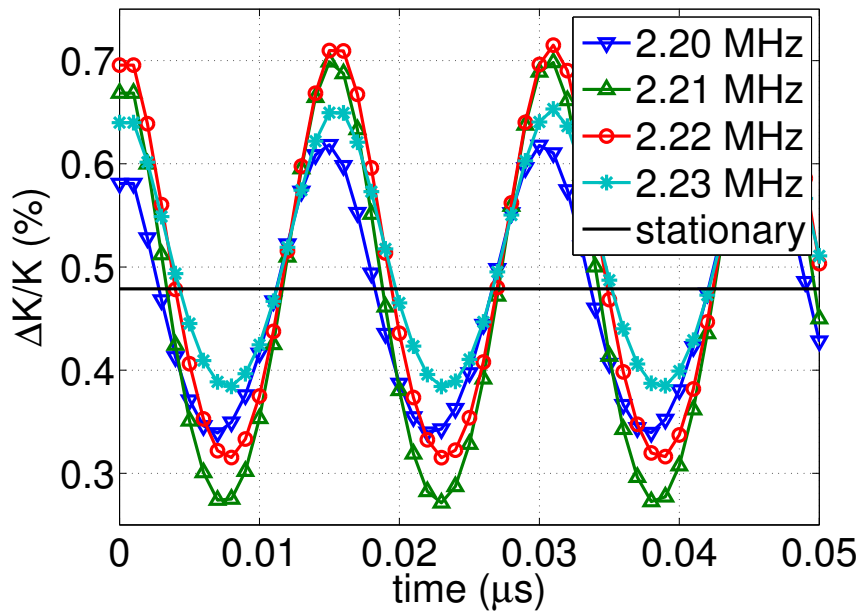}
\caption{Contact stiffness variation normalized to the baseline contact stiffness ($\Delta$K/K) for different modulation frequencies compared to the stationary case.}
\label{fig_multiplefrequencies}
\end{figure}

Furthermore, different modulation frequencies affect the contact stiffness. Fig.~\ref{fig_multiplefrequencies} shows that with modulation frequencies of \SI{2.21}{MHz} and \SI{2.22}{MHz}, the normalized contact stiffness variations are higher than for \SI{2.20}{MHz} and \SI{2.23}{MHz}. However with \SI{2.20}{MHz} and \SI{2.21}{MHz} the contrast variations are approximately symmetric with respect to the stationary case, while at \SI{2.22}{MHz} and \SI{2.23}{MHz} the contrast variations are on average higher than the stationary case. The simulation time is up to \SI{2.5}{\micro s}, however the time axis in Fig.~\ref{fig_multiplefrequencies} is limited to \SI{0.05}{\micro s}.

 Therefore, a modulation frequency of \SI{2.22}{MHz} shows approximately the same contrast as the \SI{2.21}{MHz}, but it is more efficient since the contrast minima are less pronounced than in the \SI{2.21}{MHz} case.
A non-stationary excitation has effects on the stress distribution at the tip-sample contact.  

Therefore, the modulation frequency, and thus the slow dynamics of the excitation, can be chosen to maximize the contact stiffness variation and to maximize the average of the time varying contrast with respect to the stationary case. However, the effects of such similar modulation frequencies are also emphasized here because of the choice of the carrier frequency very close to the first eigenfrequency of the system. 

In conclusion, we presented 3D ultrasound analysis and simulations of contact stiffness in US-AFM. The local variations in contact stiffness in the volume that is subjected to the ultrasound excitation influences the image contrast in US-AFM. For this reason, the effect of the ultrasound excitation needs to be included in a time-dependent approach. The presented results indicate that a static stationary approach gives an indication of the expected contrast; however, the time-dependent ultrasound approach shows that the choice of the modulation frequency, which is the slow dynamic component in the excitation, allows tailoring and optimization of the contrast.\\

This research was supported by the Early Research Program (ERP) 3D Nanomanufacturing at TNO.

\nocite{*}
\bibliography{subsurface}

\end{document}